\documentclass[twocolumn,showpacs,preprintnumbers,amsmath,amssymb,aps]{revtex4}
\usepackage{graphicx}
\usepackage{dcolumn}
\usepackage{bm}
\begin{document}
\preprint{}
\title{Neutron star matter in an effective model}
\author{T. K. Jha{\footnote {email: tkjha@phy.iitkgp.ernet.in}} ,
P. K. Raina}
\affiliation { Indian Institute of Technology, Kharagpur, India - 721302}
\author{P. K. Panda and S. K. Patra}
\affiliation { Institute of Physics, Bhubaneswar, India - 751005 }
\date{\today}
\begin{abstract}
We study the equation of state (EOS) for dense matter in the core of
the compact star with hyperons and calculate the star structure in an
effective model in the mean field approach.
With varying incompressibility and effective nucleon mass,
we analyse the resulting EOS with hyperons in beta equilibrium and
its underlying effect on the gross properties of the compact star sequences.
The results obtained in our analysis are compared with predictions of other
theoretical models and observations.
The maximum mass of the compact star lies in the range $1.21-1.96 ~M_{\odot}$
for the different EOS obtained, in the model.
\end{abstract}
\maketitle
\section{Introduction}
Dense matter studies have opened up new dimensions in understanding the nature
and behavioral aspects of nuclear matter at extremes. An ideal laboratory
for such studies can be neutron stars, which contains matter around ten
times denser than atomic nuclei. These compact stars are
believed to be made in the aftermath of type II supernova explosions
resulting from the gravitational core collapse of massive stars.
All known forces of nature i.e, strong, weak, electromagnetic
and gravitational, play key roles in the formation, evolution and the
composition of these stars. Thus the study of dense matter
not only deals with astrophysical problems such as
the evolution of neutron stars, the supernovae mechanism but also
reviews the implications from heavy-ion collisions.

Neutron stars are charge neutral, and the fact that charge neutrality
drives the stellar matter away from isospin-symmetric nuclear matter,
the study of neutron stars lends important clues in understanding
the isospin dependence of nuclear forces. Due to $\beta$-stability
conditions, neutron star is much closer to neutron matter
than the symmetric nuclear matter \cite{A1}. However, with increasing
densities, the fermi energy of the occupied baryon states reaches 
eigenenergies of other species
such as $\Lambda^{0}$(1116), $\Sigma^{-,0,+}$(1193) and $\Xi^{-,0}$(1318)
and the possibility of these hyperonic states are speculated in the
dense core of neutron stars (\cite{nkg}-\cite{mish96}). Studies on hypernuclei
experiments suggests the presence of hyperons in dense matter
such as neutron stars. Theoretically also, it has been found that the 
inclusion of hyperons in neutron star cores lowers the energy and pressure
of the system resulting in the lowering of 
the maximum mass of neutron stars, in the range of observational limits.

Various hadronic models have been applied to describe the structure of neutron
stars. Non-relativistic \cite{A2,A3} and relativistic models
(\cite{A4}-\cite{A8}) predict nearly same maximum mass of neutron star.
Relativistic models have been successfully applied to study finite nuclei
\cite{A9} and infinite nuclear matter \cite{A10} where they not only 
satisfy the properties of nuclear matter at saturation but also
the extrapolation to high density is automatically causal. Field theories such as
the non-linear $\sigma-\omega$ model \cite{boguta77} have been phenomenal in this respect.

Presently we apply an effective hadronic model to study the equation of state 
(EOS) for neutron
star matter in the mean-field type approach \cite{A10}.
Along with non-linear terms, which ensure reasonable saturation properties of
nuclear matter, the model embodies dynamical generation of the vector meson mass that
ensures a reasonable incompressibility. Therefore, one of the motivation
for the present study is to check the applicability of the model to
the study of high density matter. Secondly, the parameter sets of the model are in
accordance with recently obtained heavy-ion data \cite{daniel02}. With varying
incompressibility and effective nucleon mass the study can impart vital information
about their dependency and the underlying effect on the resulting EOS.
Also the existing knowledge on the presence of hyperons in the dense core
of these compact stars is inadequate, largely because the coupling
strength of these hyperons are unknown. So it would be interesting to
see the effect of hyperons in the dense core of neutron stars
and the predictive power of the present model in
establishing the global properties of the resulting neutron star sequences.

The outline of the paper is as follows: First we give a brief description
of the ingredients of the hadronic model that we implement in our
calculations.
After introducing the Tolman-Oppenheimer-Volkov (TOV) equations for
the static star, we present some general features of the equation of
state and then look at the gross properties of the neutron stars in our
calculations and compare our results with the observed
masses of the neutron stars, and also with predictions from some of the field-theoretical
models. We then discuss a few constraints on the neutron star mass and radius
imposed by recent estimates of the gravitational redshift in the M-R plane.
Finally we conclude with outlook on the possible extensions of the current approach.

\section{The equation of state}
We start with an effective Lagrangian generalized to include all the 
baryonic octets interacting through mesons:
\begin{widetext}
\begin{eqnarray}
\label{lag}
{\cal L}&=& \bar\psi_B~\left[ \big(i\gamma_\mu\partial^\mu
         - g_{\omega B}\gamma_\mu\omega^\mu
         - \frac{1}{2}g_{\rho B}{\vec \rho}_\mu\cdot{\vec \tau}
            \gamma^\mu\big )
         - g_{\sigma B~}~\big(\sigma + i\gamma_5
             \vec \tau\cdot\vec \pi \big)\right]~ \psi_B
\nonumber \\
&&
        + \frac{1}{2}\big(\partial_\mu\vec \pi\cdot\partial^\mu\vec\pi
        + \partial_{\mu} \sigma \partial^{\mu} \sigma\big)
        - \frac{\lambda}{4}\big(x^2 - x^2_0\big)^2
        - \frac{\lambda B}{6}\big(x^2 - x^2_0\big)^3
        - \frac{\lambda C}{8}\big(x^2 - x^2_0\big)^4
\nonumber \\
&&      - \frac{1}{4} F_{\mu\nu} F_{\mu\nu}
        + \frac{1}{2}{g_{\omega B}}^{2}x^2 \omega_{\mu}\omega^{\mu}
        - \frac {1}{4}{\vec R}_{\mu\nu}\cdot{\vec R}^{\mu\nu}
        + \frac{1}{2}m^2_{\rho}{\vec \rho}_{\mu}\cdot{\vec \rho}^{\mu}\ .
\end{eqnarray}
\end{widetext}
Here $F_{\mu\nu}\equiv\partial_{\mu}\omega_{\nu}-\partial_{\nu}
\omega_{\mu}$ and  $x^2= {\vec \pi}^2+\sigma^{2}$, $\psi_{B}$ is
the baryon spinor, ${\vec \pi}$ is the pseudoscalar-isovector pion
field, $\sigma$ is the scalar field. The subscript 
$B=n,p,\Lambda,\Sigma$ and $\Xi$, 
denotes for baryons. The terms in eqn. (1) with the subscript $'B'$ 
should be interpreted as sum over the states of all baryonic 
octets. In this model for hadronic
matter, the baryons interact via the exchange of
the $\sigma$, $\omega$ and $\rho$-meson.
The Lagrangian includes a dynamically generated mass of the
isoscalar vector field, $\omega_{\mu}$, that couples to the
conserved baryonic current $j_{\mu}=\bar{\psi_{B}}\gamma_{\mu}\psi_{B}$.
In this paper we shall be concerned only with the normal non-pion condensed
state of matter, so we take $\vec \pi=0$ and also the pion mass $m_\pi=0$.
The interaction of the scalar and the pseudoscalar mesons with the vector 
boson generate the mass through the spontaneous breaking of the chiral 
symmetry. Then the masses of the baryons, scalar and vector mesons, which are
generated through $x_0$, are respectively given by
\begin{eqnarray}
m_B = g_{\sigma B} x_0,~~ m_{\sigma} = \sqrt{2\lambda} x_0,~~
m_{\omega} = g_{\omega B} x_0\ .
\end{eqnarray}
In the above, $x_0$ is the vacuum expectation value of the $\sigma$ field,
$\lambda~=~({m_{\sigma}}^{2}-{m_{\pi}}^{2})/(2 {f_{\pi}}^{2})$, with
$m_{\pi}$, the pion mass and $f_{\pi}$ the pion decay
constant, and $g_{\omega B}$ and $g_{\sigma B}$ are the coupling constants
for the vector and scalar fields, respectively.
In the mean-field treatment we ignore the explicit role of $\pi$ mesons.

The Dirac equation for baryons is the Euler-Lagrange equation of
${\cal L}$ and is obtained as
\begin{equation}
[\gamma_\mu(p^\mu-g_{\omega B}\omega^\mu-\frac{1}{2}g_{\rho B}\vec\tau\cdot
\vec\rho^\mu)-g_{\sigma B}\sigma]\psi_B=0\ .
\end{equation}
The mass term in the above equation appears in the form $g_{\sigma B}\sigma$,
which is referred to as the effective baryon mass, $m^*_B=g_{\sigma B}\sigma$.

We will now proceed to calculate the equation of motion for the scalar field.
The scalar field dependent terms from the Lagrangian density are:
\begin{widetext}
\begin{equation}
        - \frac{\lambda}{4}\big(x^2 - x^2_0\big)^2
        - \frac{\lambda B}{6}\big(x^2 - x^2_0\big)^3
        - \frac{\lambda C}{8}\big(x^2 - x^2_0\big)^4
        - g_{\sigma B~}\bar\psi_B~\sigma ~ \psi_B
        + \frac{1}{2}{g_{\omega B}}^{2}x^2 \omega_0^2\ ,
\end{equation}
\end{widetext}
where in the mean-field limit $\omega$ = $\omega_0$.
The constant parameters $B$ and $C$ are included in the higher-order
self-interaction of the scalar field to describe
the desirable values of nuclear matter properties at saturation point.
Using equation (2) and $m^{\star}_B/m_B \equiv x/x_0 \equiv Y$, 
the above expression divided by $\lambda
x_0^4$ becomes
\begin{widetext}
\begin{equation}
-\frac{1}{4}(1-Y^2)^2 +\frac{B}{6 c_{\omega B}}(1-Y^2)^3
-\frac{C}{8c_{\omega B}^2}(1-Y^2)^4
+\frac{2 g_{\omega B}^2\omega_0^2}{2\lambda x_0^2}^2Y^2
+\frac{g_{\sigma B}Y}{\lambda x_0^3}\bar\psi_B\psi_B
\end{equation}
\end{widetext}
Differentiating with respect to $Y$, we have the equation of motion for
the scalar field including all baryons as:
\begin{widetext}
\begin{equation}
\sum_B\left[ (1-Y^2) -\frac{B}{c_{\omega B}}(1-Y^2)^2
+\frac{C}{c_{\omega B}^2}(1-Y^2)^3
+\frac{2 c_{\sigma B}c_{\omega B}\rho_B^2}{m_{B}^2Y^4}
-\frac{2 c_{\sigma B}\rho_{SB}}{m_{B} Y}\right]=0\ ,
\label{effmass}
\end{equation}
\end{widetext}
where the effective mass of the baryonic species is $m_{B}^{\star} 
\equiv Ym_{B}$
and $c_{\sigma B}\equiv  g_{\sigma B}^2/m_{\sigma}^2 $  and
$c_{\omega B} \equiv g_{\omega B}^2/m_{\omega}^2 $ are the usual
scalar and vector coupling constants respectively.
It should be noted that although the term '$\lambda$' in the
Lagrangian does not appear explicitly in eqn. (6), however the effect is
there through the mass term, following equation (2) and through $x_0$. 

For a baryon species, the scalar density ($\rho_{SB}$) and the baryon 
density ($\rho_B$) are,
\begin{equation}
\rho_{SB}= \frac{\gamma}{(2\pi)^3}\int^{k_B}_o\frac{m^*_{B} d^3k}
         {\sqrt{k^2+m_{B}^{\star 2}}},
\end{equation}
\begin{equation}
\rho_B= \frac{\gamma}{(2\pi)^3}\int^{k_B}_o d^3k,
\end{equation}
The equation of motion for the $\omega$ field is then calculated as
\begin{equation}
\omega_0=\sum_{B}\frac{ \rho_B }{g_{\omega B} x^2} \ ,
\end{equation}
The quantity $k_B$ is the Fermi momentum  for the baryon and $\gamma=2$ is the
spin degeneracy.
Similarly, the equation of motion for the $\rho-$meson
is obtained as:
\begin{equation}
\rho_{03} =\sum_{B} \frac{g_{\rho B}}{m_\rho^2} I_{3 B}\rho_{B}\ ,
\end{equation}
where $I_{3 B}$ is the 3rd-component of the isospin of each baryon species
(given in the Table II).

Traditionally, neutron stars were believed to be composed mostly of
neutrons, some of which eventually $\beta$-decay until an equilibrium
between neutron, proton and electron is reached. The respective chemical
potentials then satisfy the generic relationship,
$\mu_{p}=\mu_{n}-\mu_{e}$, among them. Along with charge
neutrality condition, $n_{p}=n_{e}$, the various particle
composition is then determined and the neutron star is believed
to be composed of neutrons, protons and electrons.
Muons come into picture when $\mu_{e}=\mu_{\mu}$, which happens roughly
around nuclear matter density, and the charge neutrality condition is altered to
$\rho_{p}=\rho_{e}+\rho_{\mu}$.
Hyperons can form in neutron star cores when the nucleon chemical potential
is large enough to compensate the mass differences between nucleon and hyperons,
which happens roughly around two times normal nuclear matter density,
when the first species of the hyperon family starts appearing.

The neutron and electron chemical potentials are constrained by the
requirements of conservation of total baryon number and the 
charge neutrality condition given by,
\begin{equation}
\sum_{B}Q_{B}\rho_{B}+\sum_{l}Q_{l}\rho_{l}=0,
\end{equation}
with $\rho_{B}$ and $\rho_{l}$ are the baryon and lepton densities respectively.
These two conditions combine to determine the appearance and concentration of
these particles in the dense core of compact objects.

A general expression may be written down for each baryonic chemical potentials ($\mu_{B}$)
in terms of these two independent chemical potentials, i.e., $\mu_{n}$ and $\mu_{e}$ as,
\begin{equation}
\mu_{B}=\mu_{n}-Q_{B}\mu_{e}
\end{equation}
where $\mu_{B}$ and $Q_{B}$ are the chemical potentials and electric charge
of the concerned baryon species.  

After achieving the solution to these conditions, one obtains the
total energy density $\varepsilon $ and pressure P for a given baryon density as:
\begin{widetext}
\begin{eqnarray}
\label{ep0}
\varepsilon
&=&
 \frac{2}{\pi^2}\int^{k_B}_0 k^{2}dk{\sqrt{k^2+m_B^{\star 2}}}
         +  \frac{m_B^2(1-Y^2)^2}{8c_{\sigma B}}
        - \frac{m_B^2 B}{12c_{\omega B}c_{\sigma B}}(1-Y^2)^3
\nonumber \\
        &+& \frac{m_B^2 C}{16c_{\omega B}^2c_{\sigma B}}(1-Y^2)^4
        + \frac{1}{2Y^2}{c_{\omega_{B}} \rho_B^2}
        +\frac{1}{2}m_{\rho}^{2}\rho_{03}^2
        + \frac{1}{\pi^{2}}\sum_{\lambda=e,\mu^{-}}\int^
           {k_\lambda}_0 k^{2}dk{\sqrt{k^2+m^2_{\lambda}}}\ ,
\end{eqnarray}
\begin{eqnarray}
P &=&
          \frac{2}{3\pi^2}\int^{k_B}_0\frac{k^{4}dk}
          {{\sqrt{k^2+m_B^{\star 2}}}}
        - \frac{m_B^2(1-Y^2)^2}{8c_{\sigma B}}
        + \frac{m_B^2 B}{12c_{\omega B}c_{\sigma B}}(1-Y^2)^3
\nonumber \\
&-&
         \frac{m_B^2 C}{16c_{\omega B}^2c_{\sigma B}}(1-Y^2)^4
        + \frac{1}{2Y^2}{c_{\omega_{B}} \rho_B^2}
        + \frac{1}{2}m_{\rho}^{2}\rho_{03}^2\
       +\frac{1}{3\pi^2}\sum_{\lambda=e,\mu^{-}}\int^{k_\lambda}_0
          \frac{k^{4}dk}{{\sqrt{k^2+m^2_{\lambda}}}}
\end{eqnarray}
\end{widetext}
As explained earlier, the terms in eqns. (13) and (14) with 
the subscript $'B'$ should be interpreted as sum over the 
states of all baryonic octets.
The meson field equations ((6), (9) and (10)) are then solved
self-consistently at a fixed baryon density to obtain the respective fields
along with the requirements of conservation of total baryon number and charge
neutrality condition given in equation (12) and the
energy and pressure is computed for the neutron star matter.
Using the computed EOS for the neutron star sequences,
we calculate the properties of neutron stars.

The equations for the structure of a relativistic spherical and
static star composed of a perfect fluid were derived from Einstein's
equations by Oppenheimer and Volkoff \cite{tov}. They are
\begin{equation}
\frac{dp}{dr}=-\frac{G}{r}\frac{\left[\varepsilon+p\right ]
\left[M+4\pi r^3 p\right ]}{(r-2 GM)},
\label{tov1}
\end{equation}
\begin{equation}
\frac{dM}{dr}= 4\pi r^2 \varepsilon,
\label{tov2}
\end{equation}
with $G$ as the gravitational constant and $M(r)$ as the enclosed
gravitational mass. We have used $c=1$.
Given an EOS, these equations can be integrated from the origin as an initial
value problem for a given choice of central energy density, $(\varepsilon_c)$.
The value of $r~(=R)$, where the pressure vanishes defines the
surface of the star.

We solve the above equations to study the structural properties of the
neutron star, using the EOS derived for the electrically charge
neutral hyperon rich dense matter.

\section{Results and discussion}
The parameter set for the present model is listed in Table-1,
which is in accordance with recently obtained heavy-ion collision data.
With varying effective masses $(m_N^{\star}=0.8-0.9~ m_{N})$ and
incompressibility ($K=210-380$ MeV), the study can give us informations on nuclear
equation of state and its effect on the properties of neutron stars.
The parameter sets satisfies the nuclear saturation properties,
$E_B$, energy per nucleon, $-16$ MeV at saturation density $0.153~ fm^{-3}$,
effective nucleon Landau mass $0.8-0.9 ~m_N$, incompressibility,
and asymmetry energy coefficient value ($\approx~ 32$ MeV), so that our
extrapolation to higher density remains meaningful.

We fix the coupling constant $c_{\rho N}$ by requiring that $a_{\rm sym}$
correspond to the empirical value, 32 $\pm$ 6 MeV\cite{moll88}.
This gives $c_{\rho N}=4.66~ \hbox{fm}^2$ for $a_{\rm sym}$=32 MeV.

\begin{table}
\caption{Parameter sets for the model.}
\vskip 0.1 in
\begin{center}
\begin{tabular}{cccccccccccc}
\hline
\hline
\multicolumn{1}{c}{set}&
\multicolumn{1}{c}{$c_{\sigma N}$}&
\multicolumn{1}{c}{$c_{\omega N}$} &
\multicolumn{1}{c}{$B$} &
\multicolumn{1}{c}{$C$} &
\multicolumn{1}{c}{$K$} &
\multicolumn{1}{c}{$m_N^{\star}/m_N$} \\
\multicolumn{1}{c}{ } &
\multicolumn{1}{c}{($fm^2$)} &
\multicolumn{1}{c}{($fm^2$)} &
\multicolumn{1}{c}{($fm^2$)} &
\multicolumn{1}{c}{($fm^4$)}&
\multicolumn{1}{c}{($MeV$)} &
\multicolumn{1}{c}{}\\
\hline
I &8.86&1.99&-12.24&-31.59&210&0.85 \\
II&6.79&1.99&-4.32&0.165&300&0.85 \\
III&5.36&1.99&1.13&22.01&380&0.85 \\
IV&8.5&2.71&-9.26&-40.73&300&0.80 \\
V&2.33&1.04&9.59&46.99&300&0.90 \\
\hline
\end{tabular}
\end{center}
\end{table}

\begin{table}
\caption{Table of baryonic octet}
\vskip 0.1 in
\begin{center}
\begin{tabular}{cccccccccc}
\hline
\hline
\multicolumn{1}{c}{B}&
\multicolumn{1}{c}{Mass} (MeV)&
\multicolumn{1}{c}{$Q_B$} &
\multicolumn{1}{c}{$I_{3}$} &\\
\hline
\hline
$p,n$              & 938  & 1,0    & 1/2,-1/2  & \\
$\Lambda^{0}$      & 1116 &   0    &    0      & \\
$\Sigma^{-,0,+}$   & 1193 & -1,0,1 & -1,0,1    & \\
$\Xi^{-,0}$        & 1318 & -1,0   & -1/2,1/2  & \\
\hline
\end{tabular}
\end{center}
\end{table}
The baryonic octet under consideration
are summarized in Table-2, with their respective masses, charge ($Q_{B}$) and
isospin ($I_{3}$). The electric charge and isospin
determine the exact conditions for each hyperon species to appear
in the matter. In the absence of any relevant data for hyperon-nucleon
or hyperon-hyperon interaction, the understanding of dense matter under
these extreme conditions heavily depends on the hypernuclei experiments.
The experiment shows bound states of $\Lambda$ in nuclear medium,
although nothing can be said about other baryon species.

In order to include hyperons, one needs to
specify the hyperon coupling strength, which is more or less unknown \cite{prak97,nkg01}.
The EOS at high density is very sensitive to the underlying hyperon couplings,
since hyperons are the majority population at high densities
and is in turn reflected in the structural properties of the compact stars.
The ratio of hyperon to nucleon couplings to the meson fields are not defined
by the ground state of nuclear matter, but are chosen on other grounds such as,

(1) Universal coupling scheme (UC):
$x_\sigma$=$x_\omega$=1, where the hyperons and nucleons couple
to the meson fields with equal strength
(2) Moszkowski coupling (MC) :
$x_\sigma$=$x_\omega$=$\sqrt{(2/3)}$ \cite{Mosz74}, which is based on the
quark sum rule approach and
(3) In our present work,
we take $x_{\sigma}$=$g_{\sigma H}$/$g_{\sigma N}$=$0.7$,
$x_{\omega}$=$g_{\omega H}$/$g_{\omega N}$=$0.783$ and $x_{\omega}$
=$x_{\rho}$, to calculate the EOS for the neutron star matter 
and gross properties of neutron stars. Here, 
binding of $\Lambda^{0}$ in nuclear matter:
$(B/A)_{\Lambda}$=$x_{\omega}g_{\omega}\omega_{0}+m^{*}_{\Lambda}
-m_{\Lambda}\approx-30$ MeV.
However prescription (3) restricts the equation of state of
neutron star matter following the constraint of $\Lambda^{0}$ binding in nuclear matter.
The choice of $x_\sigma < 0.72$ has been emphasized \cite{glen91} and also from studies
based on hypernuclear levels \cite{rufa90}, the choice ($x_\sigma < 0.9$)
is bounded from above.

\begin{figure}[ht]
\begin{center}
\includegraphics[width=8cm,height=10cm,angle=-90]{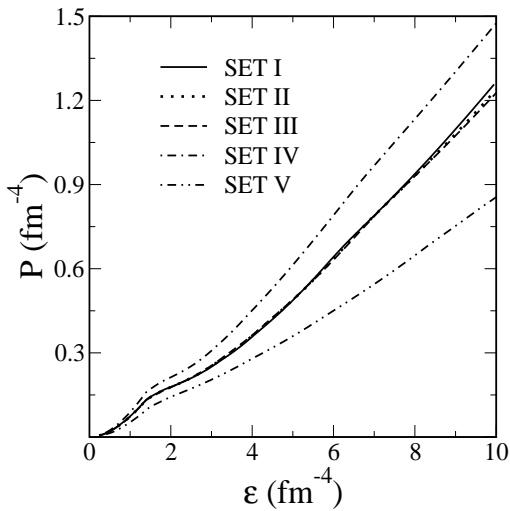}
\end{center}
\caption
{Equation of state ($P$ vs $\varepsilon$) of neutron star matter with hyperons
for the parameter sets listed in Table 1.}
\end{figure}

The nucleon effective mass and incompressibility strongly influence
the EOS of neutron-rich and neutron star matter.
Figure 1 displays the equation of state for the five parameter sets.
From the figure, it is to be noted that parameter set I, II, and III
(with same nucleon effective mass but different incompressibility) follows
similar trend upto ten times normal nuclear matter density, whereas sets
IV and V represents the stiff and soft character of the EOS respectively, which
can be attributed to their different effective mass values.
Thus the difference in incompressibility does not seem to bother
much to the resulting equation of state, whereas the difference in
effective mass appears to be prominent. However, for all
the cases, we find the dip in the curve at $\varepsilon$ $\approx$ 1.5-2 $fm^{-4}$,
which is the signature of the appearance of first members of
the hyperons family namely $\Lambda^{0}$ and $\Sigma^{-}$ states.

Similarly, successive appearance
of each of the hyperon species contributes to the softening of
the EOS. Similar feature is noticeable in most of the other relativistic
field theoretical models \cite{mishra2002}. Here, it can be seen that
a higher effective mass value (Set V) results in a softer EOS and vice-versa,
whereas difference in incompressibility is visible at higher densities only.

\begin{figure}[ht]
\begin{center}
\includegraphics[width=8cm,height=10cm,angle=-90]{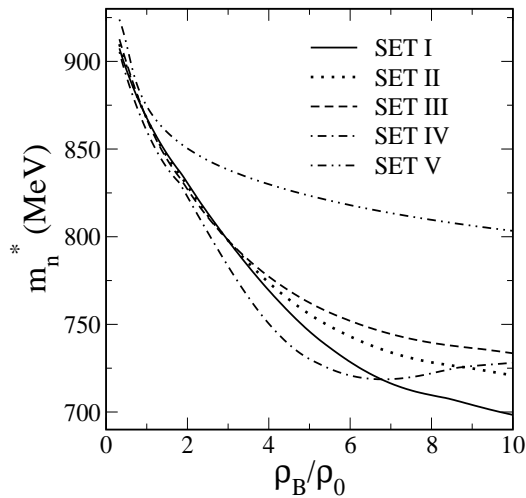}
\end{center}
\caption
{ Effective nucleon mass as a function of baryon density in the neutron star
matter upto $10\rho_{0}$.
}
\end{figure}

\begin{figure}[ht]
\begin{center}
\includegraphics[width=8cm,height=10cm,angle=-90]{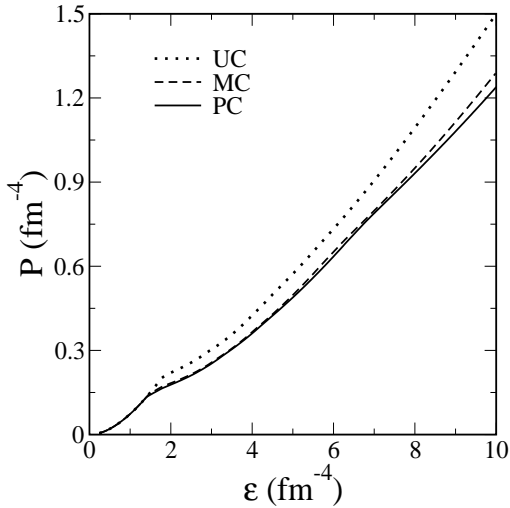}
\end{center}
\caption
{Equation of state for SET II corresponding to different hyperon
coupling schemes, as described in the text. }
\end{figure}

The nucleon effective mass ($m_{N}^{\star} \equiv Ym_{N}$) as a function of baryon
density upto ten times normal nuclear matter density is displayed in figure 2.
In case of set I, II and III, the nucleon effective mass follows similar trend,
where the nucleon sheds around 22-25 $\%$ of its mass in the matter upto $10\rho_{0}$.
Set IV follows the same trend till 6-7$\rho_{0}$  but
then the mass increases slowly, the strong repulsive
component is responsible for the feature. In case of set V, the nucleon sheds
only 15 $\%$ of its mass upto ten times nuclear density, but the
decrease is rather much slower after a steep decrease till 2$\rho_{0}$.
This gradual decrease is as a result of strong scalar component in set V.

To show the sensitivity of EOS to that of the hyperon couplings, we compare in
figure 3, EOS corresponding to three different coupling strength for parameter set II.
>From the figure, it can be seen that Universal coupling predicts a stiff EOS
followed by Moszkowski coupling. The coupling strength employed here
for the present calculation (PC) is closer to that of MC and is the softest
among the three prescriptions. Thus it is conclusive that weaker hyperon coupling
leads to softer equation of state because of the underlying weak repulsion in the matter.
Similar feature has been noticed in works by Glendening \cite{Glend89}
and Ellis et. al \cite{Ellis91}.

\begin{figure}[ht]
\begin{center}
\includegraphics[width=8cm,height=10cm,angle=-90]{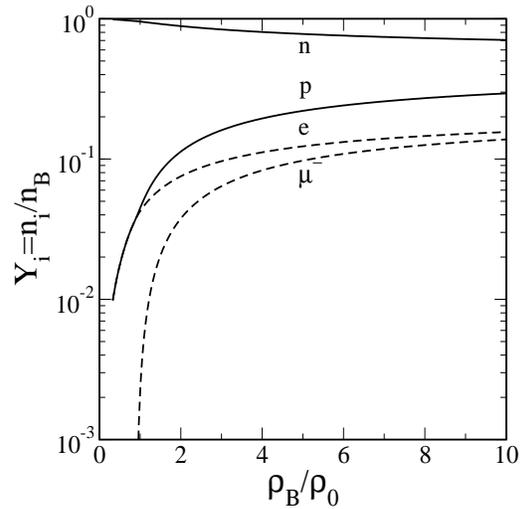}
\end{center}
\caption
{Relative Particle Population for $\beta$-equiliberated matter
without hyperons.  }
\end{figure}

For the sake of completeness we plot in figure 4 the respective particle population
of $n$, $p$, $e$ and $\mu^{-}$ matter in beta equilibrium upto $10\rho_{0}$.
Muons appear when the chemical potential of the electrons
exceeds the rest mass of the muons (106 MeV), which happens roughly at around normal
nuclear matter density and becomes one of the particle species in the composition.
Consequently the proton fraction increases with appearance of muons in the medium to
maintain charge neutrality of the matter.
The proton fraction and the electron chemical potential have been found to be
important in assessing the cooling rates of neutron stars \cite{Peth92},
and the possibility of Kaon condensation in neutron star interiors
\cite{Bro94,Pan95}. We refrain ourselves from further details in this direction.

\begin{figure}[ht]
\begin{center}
\includegraphics[width=8cm,height=10cm,angle=-90]{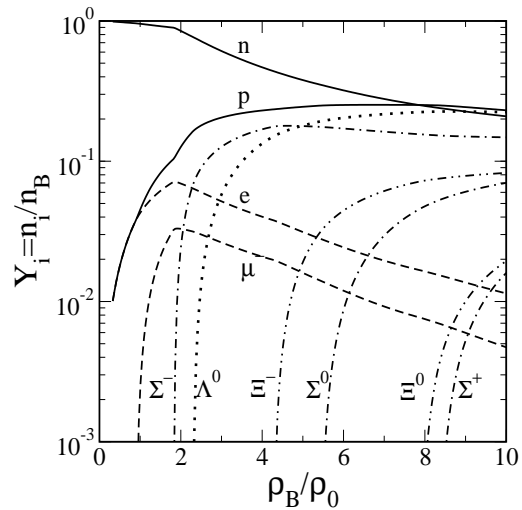}
\end{center}
\caption
{
Relative particle population for neutron star matter with hyperons for
parameter Set I (K=210 MeV, $m_N^{\star}$ =0.85 $m_N$).
}
\end{figure}

Figure 5, 6 and 7 displays the relative particle composition of neutron star
matter for parameter sets I, II and III with all baryon octets in equilibrium
rendering a charge neutral hyperon rich matter. From the plots, it is noteworthy
that the difference in the incompressibilities doesn't manifest in the particle
composition of the matter very much.

\begin{figure}[ht]
\begin{center}
\includegraphics[width=8cm,height=10cm,angle=-90]{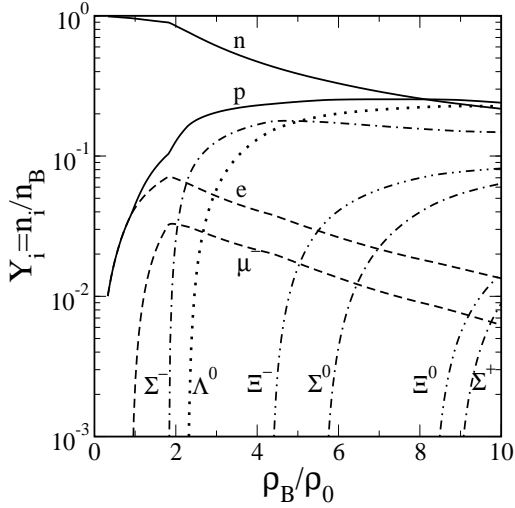}
\end{center}
\caption
{
Relative particle population for neutron star matter with hyperons for
Set II (K=300 MeV, $m_N^{\star}$ =0.85 $m_N$).
}
\end{figure}

\begin{figure}[ht]
\begin{center}
\includegraphics[width=8cm,height=10cm,angle=-90]{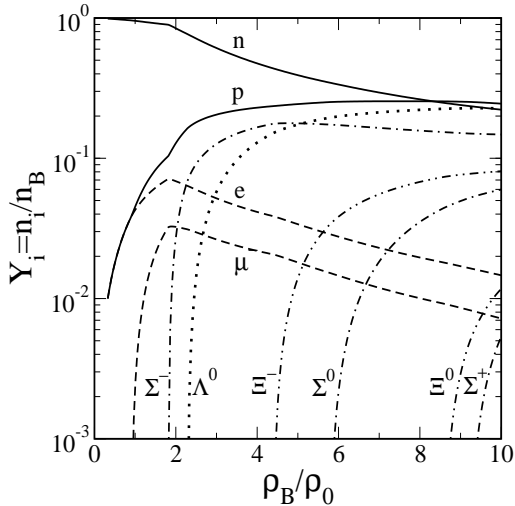}
\end{center}
\caption
{
Relative particle population for neutron star matter with hyperons for
Set III (K=380 MeV, $m_N^{\star}$ =0.85 $m_N$).
}
\end{figure}

In all three cases, the hyperons start appearing at around $2\rho_{0}$, where
$\Sigma^{-}$ appears first, closely followed by $\Lambda^{0}$. However the former
gets saturated because of isospin dependent forces and $\Lambda^{0}$
exceeds $\Sigma^{-}$ roughly at about 4$\rho_{0}$. $\Lambda^{0}$ density further
increases with increasing baryon density and is in fact, the dominant
particle in the matter composition along with the nucleons. At higher densities
other baryon thresholds are attained and they also start appearing.
Only noticeable difference in the three cases are the relative population
density of the higher octets namely $\Xi^{0}$ and $\Sigma^{+}$ states.
With increasing incompressibility the density at which they appear is pushed further
and the population of these states decreases accordingly, although not appreciably.

\begin{figure}[ht]
\begin{center}
\includegraphics[width=8cm,height=10cm,angle=-90]{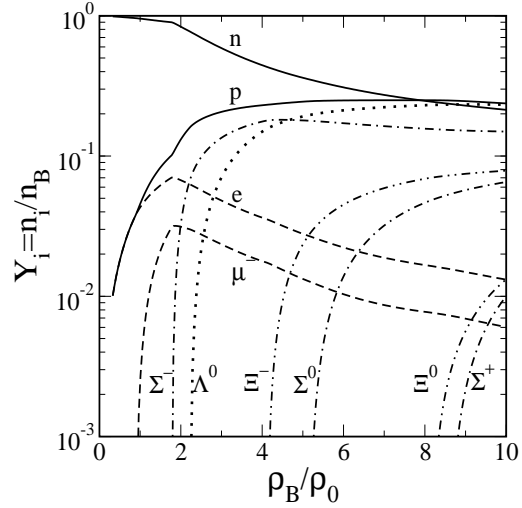}
\end{center}
\caption
{
Relative particle population for the neutron star matter with hyperons for
Set IV (K=300 MeV, $m_N^{\star}$ =0.80 $m_N$).
}
\end{figure}

\begin{figure}[ht]
\begin{center}
\includegraphics[width=8cm,height=10cm,angle=-90]{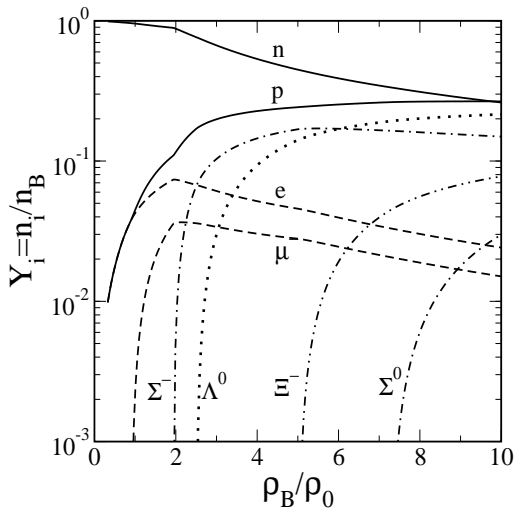}
\end{center}
\caption
{
Relative particle population for the neutron star matter with hyperons for
Set V (K=300 MeV, $m_N^{\star}$ =0.90 $m_N$).
}
\end{figure}

Similarly figure 8 and figure 9 display the relative particle population for parameter set IV and
V respectively. These two parameters differ in their effective nucleon mass, $m_N^{*}=$0.85 $m_N$
for set IV and $m_N^{*}=$0.90 $m_N$ for set V but with same incompressibility ($K$=300 MeV).
The difference in effective mass is very much pronounced and is reflected
in the respective particle composition of the neutron star matter.
The matter composition in case of set IV is more or less similar to
first three sets, except that the deleptonisation occurs rather differently. Set V
don't predict the presence of $\Sigma^{+}$ state even upto 10$\rho_{0}$.
In fact, all the hyperon species seem to appear at relatively higher density than
in case of set IV. For example, $\Sigma^{0}$ state appears around 7.5$\rho_{0}$ in case
of set V whereas set IV predicts that same at around 5.3$\rho_{0}$.
Due to the high effective mass value of set V, the hyperons don't seem to enjoy the
previlage of being the dominant species in par with nucleons in the matter as in
case of other parameter sets. Even at $10\rho_{0}$ nucleons comprise $\approx$ 52$\%$
of the total matter.

\begin{figure}[ht]
\begin{center}
\includegraphics[width=8cm,height=10cm,angle=-90]{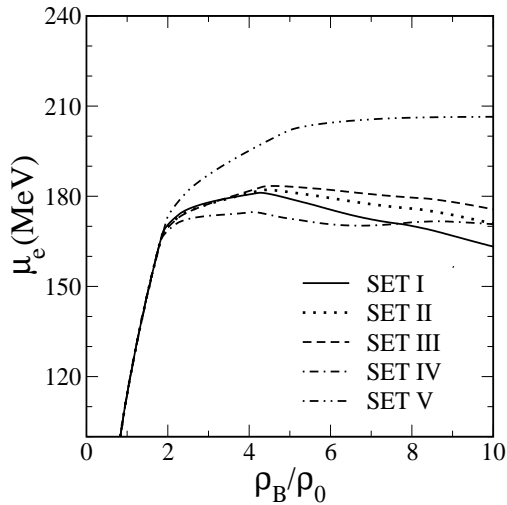}
\end{center}
\caption{Electron chemical potential as a function of baryon density upto
$10~\rho_0$.}
\end{figure}

However for all the five cases, the negatively
charged particles are found to be highly favored species in dense matter as evident
from their respective order of appearance. At higher density, it can be seen that
hyperons forms a sizable population in neutron star matter.
They clearly softens the equation of state at high densities.
The two independent chemical potentials,
$\mu_{n}$ and $\mu_{e}$ along with the charge neutrality condition decides
the particle composition. The electron chemical potentials for the five sets are
displayed in Figure 10. It can be seen that $\mu_{e}$ follows similar
pattern for parameter sets I, II and III, but in case of set IV, it
follows similar pattern until $\approx$ 6$\rho_{0}$ after which it increases again,
whereas for set V, it saturates at $\approx$ 6$\rho_{0}$ and thereby remains constant.
For all the sets at $\approx 2~\rho_0$, we see a sharp turn in the
electron potential,
where the first charged hyperon species, $\Sigma^-$ appears. At that point
$\mu_e$ compensates the mass difference between $\Sigma^{-}$ and $\Lambda^0$
thereby triggering the appearance of the former.
Leptons primarily maintain charge neutrality of the matter,
and since set V does not predict any charged hyperon species after
$\approx 6~\rho_0$,
the electron potential remains constant thereafter.

\begin{figure}[ht]
\begin{center}
\includegraphics[width=8cm,height=10cm,angle=-90]{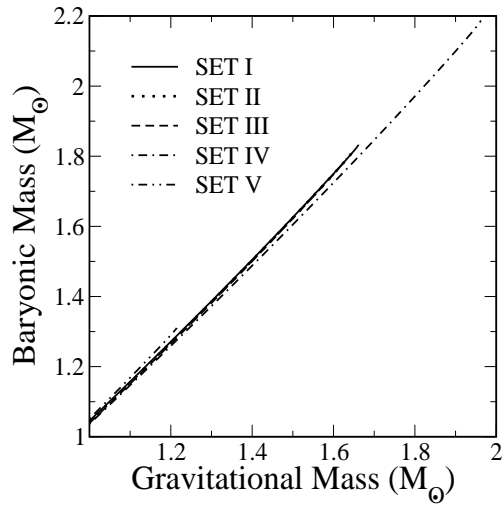}
\end{center}
\caption
{
Baryonic mass ($M_{\odot}$) of the star as a function of Maximum mass ($M_{\odot}$)
for the five sets.
}
%\label{}
\end{figure}

As stated, the properties of neutron star is unique to the EOS considered.
Using these EOS, we now calculate some of the global
properties of neutron star by solving the TOV equation.
Figure 11 shows the maximum baryonic mass $M_{b}$ ($M_{\odot}$) obtained as a
function of star mass for the five parameter sets. The curves for set I, II and III
coincides with each other, whereas set IV and V are distinctly apart, the
reason can be attributed to their different effective mass values.
However the baryonic mass always exceeds the gravitational mass, which is
typical of compact objects. The difference between the two is defined as
the gravitational binding of the star.
The baryonic masses obtained for set I, II and III are 1.83$M_{\odot}$,
1.81$M_{\odot}$ and 1.79$M_{\odot}$ respectively.
Whereas sets IV (stiff) and V (soft) EOS represents the two extremes among
all the parameter sets. The corresponding baryon masses obtained are
2.18$M_{\odot}$ and 1.31$M_{\odot}$.

\begin{figure}[ht]
\begin{center}
\includegraphics[width=8cm,height=10cm,angle=-90]{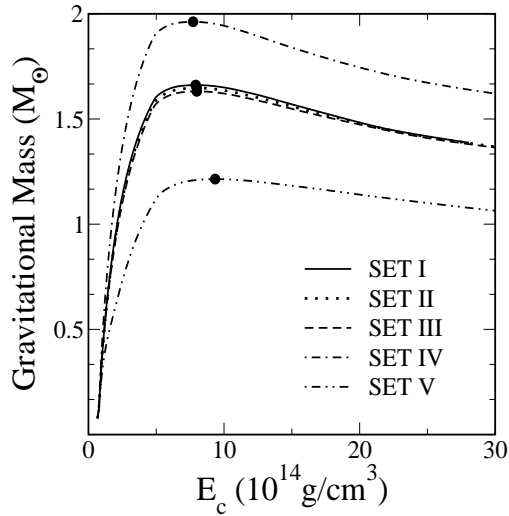}
\end{center}
\caption
{
Maximum mass of the neutron star sequences as a
function of central density of the star (in $10^{14} g cm^{-3}$).
}
\end{figure}

\begin{figure}[ht]
\begin{center}
\includegraphics[width=8cm,height=10cm,angle=-90]{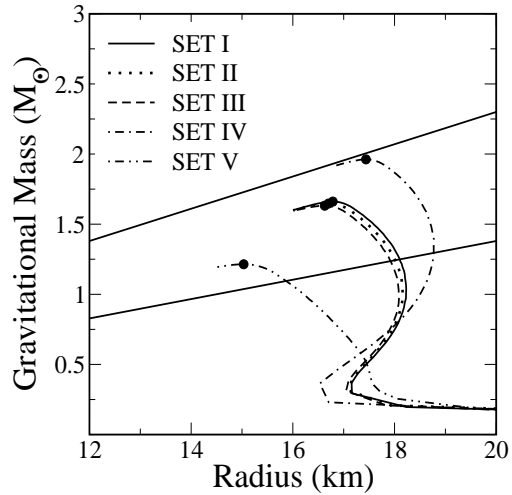}
\end{center}
\caption
{
Maximum mass of the neutron star (in solar mass) as a
function of radius (in Km) for the five Sets. The two solid
curves corresponds to $M/R=0.069$ and $M/R=0.115$.
(The solid circles represent the values at maximum mass.)
}
\end{figure}

\begin{figure}[ht]
\begin{center}
\includegraphics[width=8cm,height=10cm,angle=-90]{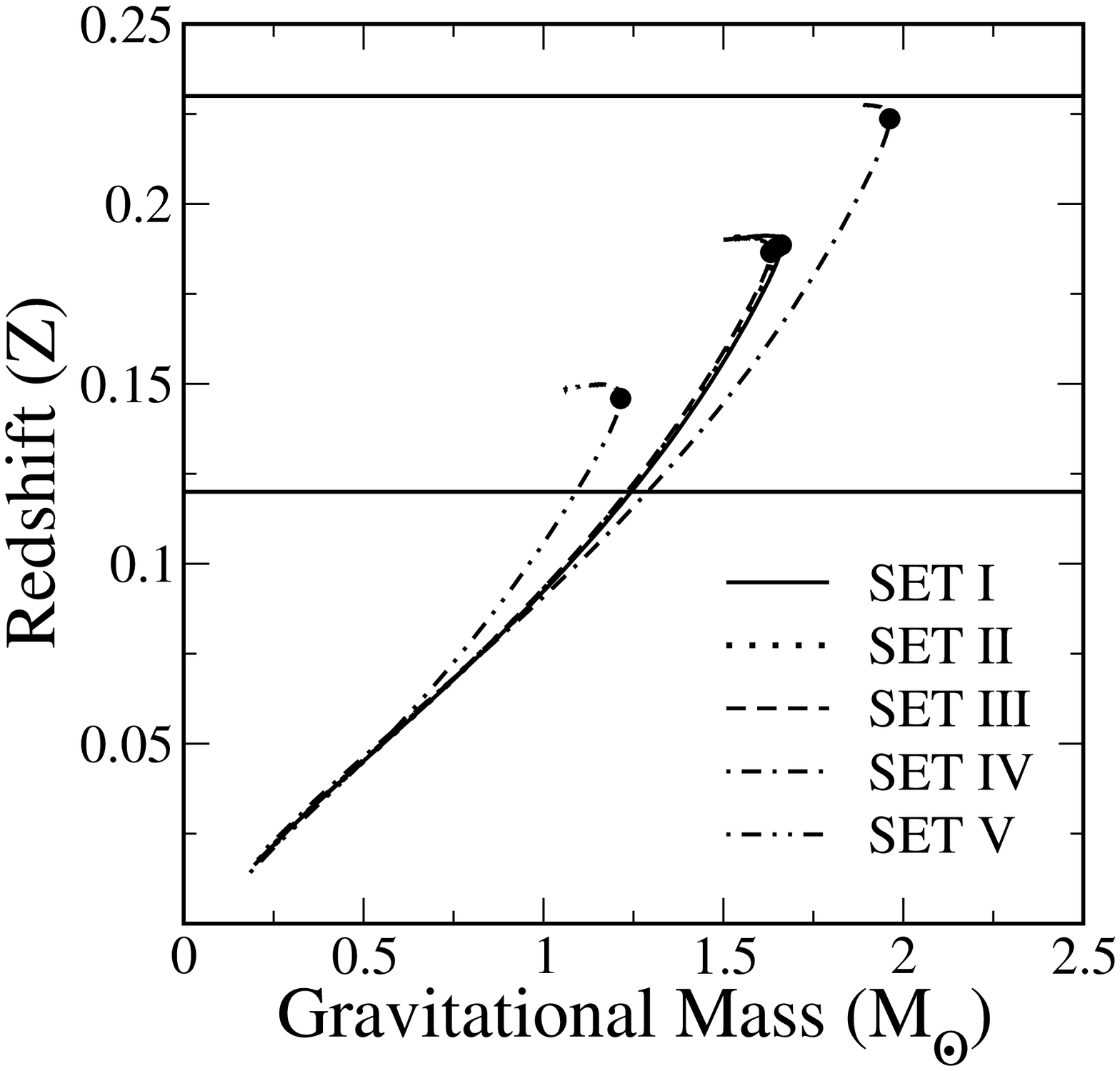}
\end{center}
\caption
{
Gravitational Redshift (Z) as a function of Maximum mass of the neutron star
for the five parameter sets.
(The solid circles represent the values at maximum mass).
The area between solid horizontal lines represents the
redshift values $Z=(0.12-0.23)$ \cite{san02}.
}
%\label{}
\end{figure}

Gravitational mass of the neutron star as a function of central density of
the star is plotted in Figure 12.
Stable neutron star configurations are the regions where $\frac{dM}{d\varepsilon_c}> 0$.
Beyond the maximum mass, gravity overcomes and results in the collapse of the star.
Set I, II and III which vary in incompressibilities predicts almost same
central density $\approx$ 7.9 $\times$ $10^{14} g cm^{-3}$ for the star at maximum mass
denoted by filled circles in the plot. The maximum mass obtained are 1.66$M_{\odot}$,
1.65$M_{\odot}$ and 1.63$M_{\odot}$ for set I, II and III respectively.
Set IV and V predicts the maximum mass to be 1.96$M_{\odot}$ and 1.21$M_{\odot}$
respectively with corresponding central densities 7.7$\times$ $10^{14} g cm^{-3}$ and
9.3$\times$ $10^{14} g cm^{-3}$ respectively.
Recent observations of neutron star masses like $M_{J0751\pm1807}$
=2.1$\pm$0.2 $M_{\odot}$\cite{ns01}, $M_{4U 1636\pm536}$=2.0$\pm$0.1 $M_{\odot}$\cite{ns02},
$M_{Vela X-1}$=1.86$\pm$0.16 $M_{\odot}$\cite{ns03} and
$M_{Vela X-2}$=1.78$\pm$0.23 $M_{\odot}$\cite{ns04,ns05} predicts massive stars.
Our results agrees remarkably with these observed masses except for set V.

Figure 13 displays the maximum mass of the neutron star as a function of
the star radius. In order to calculate the radius, we included the results
of Baym, Pethick and Sutherland \cite{bps} EOS at low baryonic densities.
The radius predicted for the sets I, II, and III are $\approx$ 16.7 km,
whereas for set IV and V, it comes out to be 17.43 and 15 km respectively.
It is to be noted that in the relativistic regime, the maximum masses obtained by
the non-linear walecka model (NLWM) and the quark-meson coupling model
\cite{panda05} are 1.90 $M_{\odot}$ and 1.98 $M_{\odot}$ respectively. 
The masses obtained in our calculations are in fair agreement 
with these calculations.
In the relativistic mean field approach the properties of neutron
star was studied \cite{A6} where it was pointed out that a bigger effective
nucleon mass results in a low mass star but with larger radius. Our results
lead to the same interpretation.

However because wide range of masses and radius of neutron star being placed
by different models, it is therefore important to impose constraints that
can put stringent condition in the M-R plane.
Constraints on the mass-radius plane can be obtained from accurate measurements
of the gravitational redshift of spectral lines produced in neutron star photospheres.
Measuring M/R is particularly important to constrain the EOS of dense matter.

Recently a constraint to M-R plane was reported \cite{san02} based on the
observation of two absorption features in the source spectrum of the IE 1207.4-5209
neutron star, which limits M/R=(0.069 - 0.115) $M_{\odot}$/km.
The region enclosed in figure 13 by two solid lines denotes the area enclosed
in accordance with the observed range. All the parameter set of the present model
satisfies the criterian very well.
Another important aspect of compact objects is the observed gravitational redshift,
which is given by
\begin{eqnarray}
Z=\frac{1}{\sqrt{1-2GM/Rc^{2}}}
\end{eqnarray}

\begin{table}
\caption{Properties of Neutron star as predicted by the model}
\vskip 0.1 in
\begin{center}
\begin{tabular}{cccccccccccc}
\hline
\hline
\multicolumn{1}{c}{SET} &
\multicolumn{1}{c}{$M(M_{\odot}$}) &
\multicolumn{1}{c}{$E_{c}$($10^{14}gcm^{-3}$)} &
\multicolumn{1}{c}{$R$}($Km$) &
\multicolumn{1}{c}{$M_{b}$($M_{\odot}$)} &
\multicolumn{1}{c}{$Z$} \\
\hline
\hline
I       &1.66  &7.90  &16.78  &1.83 &0.19\\
II      &1.65  &7.99  &16.70  &1.81 &0.19\\
III     &1.63  &7.99  &16.62  &1.79 &0.19\\
IV      &1.96  &7.72  &17.44  &2.18 &0.22\\
V       &1.21  &9.34  &15.03  &1.31 &0.15\\
\hline
\end{tabular}
\end{center}
\end{table}

The gravitational redshift interpreted by the $M/R$ ratio comes out to be in the
range $Z$=0.12-0.23, which is plotted in figure 14.
For Set I, II and III, the redshift is nearly same because redshift primarily depends
on the mass to radius ratio of the star, which in case of first three sets is nearly same.

For all the parameter sets, the redshift obtained at maximum mass lies in the range
$(0.15-0.22)$, which corresponds to $R/M=(8.8-14.2)$ km/$M_{\odot}$.
Our calculations predicts $R/M$ in the range $(8.90-12.40)$ km/$M_{\odot}$,
which is consistent with the observed value.
The predictive power of the model is evident from figure 14, where we
compare the gravitational redshift as a function of the star mass
for the five parameter sets.
The overall results of our calculation are presented in Table 3.

\section{Summary and outlook}

We studied the equation of state of high density matter in an effective
model and calculated the gross properties for neutron stars like mass, radius,
central density and redshift. We analysed five set of parameters with incompressibility values
$K$=210, 300 and 380 MeV and effective masses $m^{\star}$
=0.80, 0.85 and 0.90 $m_n$, that satisfies the nuclear matter
saturation properties. The results are then compared with some recent observations
and also a few field theoretical models. It was found that the difference in
nuclear incompressibility is not much reflected in either equation of state
or neutron star properties, but nucleon effective masses were quite decisive.
At maximum mass, the central density of the star for sets I, II, III and IV
was found to be $\approx 3~\rho_{nm}$ (nuclear matter density) but
for set V, it was found to be $\approx 3.5~\rho_{nm}$, which has the
highest effective mass value. Similarly the maximum mass
obtained for the the five EOS lies in the range 1.21-1.96$M_{\odot}$.
Set V, which is softest among all parameter sets, predicts lowest
maximum mass 1.21$M_{\odot}$, whereas set IV (stiff) predicts the maximum
mass to be 1.96$M_{\odot}$ and also is the star with the largest radius.
The difference in maximum mass and radius of
the star in case of set I, II and III is negligible, and so the predicted
redshift comes out nearly same, whereas set IV and V presents the
two extremes in overall properties, which is the reflection of their
different effective mass values. Overall, mass predicted by all the
parameter sets agree well with most of the theoretical work and
observational limits.

The results were also found to be in good agreement with recently imposed
constrains on neutron star properties in the M--R plane, and the
redshift interpreted therein. Further, the precise measurements of mass of
both neutron stars in case of PSR B1913+16 \cite{Tay89}, PSR B1534+12 \cite{Wol91}
and PSR B2127+11C \cite{Dei96} are available which can put constrains on the
nuclear equation of state. Masses of neutron stars in X-ray pulsars
are also consistent with these values, although are measured less accurately.
In case of radii, the values are still unknown, however some estimates are
expected in a few years, which would further constrain the EOS of neutron
star in the M-R plane.
In future, we intend to study the effect of rotation
to neutron star structure and also the phase transition aspects in the model.
It is worth mentioning that the density-dependent meson-nucleon couplings
is very much successful in non-linear Walecka model\cite{ring02}
and similar work in this direction would be interesting.

\begin{acknowledgments}

One of us TKJ would like to thank facilities and hospitality provided by
Institute of Physics, Bhubaneswar where a major part of the work was done.
This work was supported by R/P, under DAE-BRNS, grant no 2003/37/14/BRNS/669.

\end{acknowledgments}


\begin{thebibliography}{99}
\bibitem{A1}S.L. Shapiro and S.A. Teukolski, {\it Black holes, white dwarfs,
and Neutron stars} (Wiley, New York, 1983).
\bibitem{nkg}N.K. Glendening, Phys. Lett. {\bf B114}, 392 (1982);
N. K. Glendening, Astrophys. J. {\bf 293}, 470 (1985);
N. K. Glendening, Z. Phys. {\bf A 326}, 57 (1987).
\bibitem{prak97} M. Prakash, I. Bombaci, M. Prakash, P.J. Ellis, J.M.
Lattimer and R. Knorren, Phys. Rep. {\bf 280}, 1 (1997).
\bibitem{mish96}J. Schaffner-Beilich and I.N. Mishustin, Phys. Rev. {\bf C 53}, 1416 (1996).
\bibitem{A2}A. Akmal, V.R. Pandharipande and D.G. Ravenhall, Phys. Rev.
{\bf C 58}, 1804 (1998).
\bibitem{A3}R.B. Wiringa, V. Fiks and A. Fabrocini, Phys. Rev. 
{\bf C 38}, 1010 (1988).
\bibitem{A4}J.D. Walecka, Ann. Phys. {\bf 83}, 491 (1974).
\bibitem{A5}A. Lang, B. Blattel, W. Cassing, V. Koch, U. Mosel and K.
Weber, Z. Phys. {\bf A 340}, 207 (1991).
\bibitem{A6}N. K. Glendening, F. Weber and S.A. Moszkowski, Phys.
Rev. {\bf C 45}, 844 (1992).
\bibitem{A8}H. Heiselberg and M. Hjorth-Jensen, Astro. J. Lett. 
{\bf 525}, L45 (1999).
\bibitem{A9}T. Sil, S.K. Patra, B.K. Sharma, M. Centelles and X. Vi\~nas,
Phys. Rev. {\bf C 69}, 044315 (2004); M. Del Estal, M. Centelles, X. Vi\~nas,
and S.K. Patra, Phys. Rev. {\bf C 63}, 044321 (2001); S. K. Patra, 
M. Del Estal, M. Centelles and X. Vi\~nas, Phys. Rev. {\bf C 63}, 024311 (2001).
\bibitem{A10}P. Arumugam, B.K. Sharma, P.K. Sahu, S. K. Patra, T. Sil, 
M. Centelles and X. Vi\~nas, Phys. Lett. {\bf B 601}, 51 (2004); 
P.K. Panda, A. Mishra, J.M. Eisenberg and W. Greiner, Phys. Rev. {\bf C56},
3134 (1997).
\bibitem{boguta77}J. Boguta and A.R. Bodmer, Nucl. Phys. 
{\bf A 292}, 413 (1977).
%\bibitem{bonche81}P. Bonche and D. Vautherin, Nucl. Phys. 
%{\bf A 372}, 496 (1981).
%\bibitem{serot86}B.D. Serot and J.D. Walecka, Adv. Nucl. Phys. {\bf 16}.
\bibitem{daniel02}P. Danielewicz, R. Lacey, W.G. Lynch, Science 
{\bf 298}, 1592 (2002).
\bibitem{tov}J.R. Oppenheimer and G.M. Volkoff, Phys. Rev {\bf55},
374 (1939); R.C. Tolman, Phys. Rev {\bf55}, 364 (1939).
\bibitem{moll88}P. M\"oller, W.D. Myers, W.J. Swiatecki and J. Treiner,
At. Data Nucl. Data Tables {\bf 39}, 225 (1988).
\bibitem{nkg01}N.K. Glendening Phys. Rev. {\bf C 64}, 025801 (2001).
\bibitem{Mosz74}S.A. Moszkowski, Phys. Rev. {\bf D 9}, 1613 (1974).
\bibitem{glen91}N.K. Glendening and S.A. Moszkowski, Phys. Rev. Lett. 
{\bf 67}, 2414 (1991).
\bibitem{rufa90}M. Rufa, J. Schaffner, J. Maruhn, H. Stocker, W. Greiner 
and P. G. Reinhard, Phys. Rev. {\bf C 42}, 2469 (1990).
\bibitem{mishra2002}A. Mishra, P.K. Panda and W. Greiner, J. Phys. 
{\bf G 28}, 67 (2002).
\bibitem{Glend89}N.K. Glendening, Nucl. Phys. {\bf A 493}, 521 (1989).
\bibitem{Ellis91}J. Ellis, J.I. Kapusta and K.A. Olive, Nucl. Phys. 
{\bf B 348}, 345 (1991).
\bibitem{Peth92}C.J. Pethick, Rev. Mod. Phys. {\bf64}, 1133 (1992).
\bibitem{Bro94}G.E. Brown, C.H. Lee, M. Rho and V. Thorsson, Nucl Phys.
{\bf  A 567}, 937 (1994).
\bibitem{Pan95}V.R. Pandharipande, C.J. Pethick and V. Thorsson, Rev. Lett.
{\bf75}, 4567 (1995).
\bibitem{ns01}D.J. Nice, E.M. Spalver, I.H. Stairs, O. Loehmer, A. Jessner, 
M. Kramer and J.M. Cordes, Astrophys. J. {\bf 634} 1242 (2005).
\bibitem{ns02} D. Barret, J.F. Olive and M.C. Miller, {\it astro-ph/0605486}.
\bibitem{ns03} O. Barziv, L. Kaper, M.H. van Kerkwijk, J.H. Telting and J. van Paradijs,
Astron. \& Astrophys. {\bf 377} 925 (2001).
\bibitem{ns04}J. Casares, P.A. Charles and E. Kuulkers, Astro. J. 
{\bf 493} L39 (1998).
\bibitem{ns05}J.A. Orosz and E. Kuulkers, Mon. Not. R. Astron. Soc. 
{\bf 305} 132 (1999).
\bibitem{bps}G. Baym, C. Pethick and P. Sutherland, Astrophys. J. 
{\bf 170}, 299 (1971).
\bibitem{panda05}P.K. Panda, D.P. Menezes, C. Provid\^ncia, Phys.Rev. {\bf C69},
025207 (2004); P.K. Panda, D.P. Menezes, C. Provid\^ncia, Phys.Rev. {\bf C69},
058801 (2004); D.P. Menezes, P.K. Panda and C. Provid\^encia,
Phys. Rev. {\bf C 72}, 035802 (2005).
\bibitem{san02}D. Sanyal, G.G. Pavlov, V.E. Zavlin and 
M.A. Teter, Astrophys. J. {\bf 574} L61 (2002).
\bibitem{Tay89}J.H. Taylor and J.M. Weisberg, Astrophys. J. {\bf345}, 434 (1989).
\bibitem{Wol91}A. Wolsczan, Nature {\bf350}, 688 (1991).
\bibitem{Dei96}W.T.S. Deich, S.R. Kulkarni in {\it Compact Stars in Binaries},
J. van Paradijs, E.P.J. van den Heuvel and E. Kuulkers eds., 
Dordrecht, Kluwer (1996).
\bibitem{ring02} T. Niksic, D. Vretenar, P. Finelli and P. Ring,
Phys. Rev. {\bf C66}, 024306 (2002).
\end{thebibliography}
\end{document}